\documentclass[prd,showkeys,floatfix,nofootinbib,
               preprint,12pt,
               tightenlines,fleqn]{revtex4} 

\usepackage{amsmath,amssymb,revsymb,graphicx,dcolumn}
\usepackage{leftidx}

\newcommand{\beq}{\begin{equation}}
\newcommand{\eeq}{\end{equation}}
\newcommand{\beqa}{\begin{eqnarray}}
\newcommand{\eeqa}{\end{eqnarray}}
\newcommand{\bsubeqs}{\begin{subequations}}
\newcommand{\esubeqs}{\end{subequations}}

\begin{document}
\title[]
      {Non-unitarity or hidden observables?\vspace*{5mm}}
\author{Slava Emelyanov}
\email{viacheslav.emelyanov@physik.uni-muenchen.de}
\affiliation{Arnold Sommerfeld Center for
Theoretical Physics,\\
Ludwig Maximilian University (LMU),\\
80333 Munich, Germany\\}

\begin{abstract}
\vspace*{2.5mm}\noindent
A free hermitian conformal field theory is considered in Minkowski, de Sitter and anti-de Sitter spacetimes. The first 
part of the paper studies spacetime inversion and conformal inversion, wherein their role in the field quantization is 
elucidated in those spaces. The second part of the paper is concerned with the non-unitary evolution of detector's 
state. Several examples of such processes are provided with a clarification of how the unitarity is preserved with still 
having well-known thermal effects.
\end{abstract}



\maketitle

\section{Introduction}

Among symmetries of the maximally symmetric spacetimes there are those which are not generated by the algebra 
of the conformal Killing vectors $L_{AB} \in \mathfrak{so}(2,4)$.\footnote{The conformal Lie algebra $\mathfrak{so}(2,4)$ 
is generated by $L_{AB} = \xi_A\partial_B - \xi_B\partial_A$, where $L_{AB}$ satisfy 
$\big[L_{AB},L_{CD}\big] = \eta_{AC}L_{DB} + \eta_{AD}L_{BC} + \eta_{BC}L_{AD} + \eta_{BD}L_{CA}$,
$\eta_{AB} = \text{diag}(+,-,-,-,-,+)$ and $A,B = 0,..,5$. Minkowski, dS and AdS spaces are realized as 4D subspaces in this
6D one. By definition $\hat{L}_{AB} \equiv - i L_{AB}$ below.} These discrete symmetries 
play a significant role in the quantum theory. In particular, the spacetime inversion $\mathfrak{I}$ in Minkowski space 
and the charge conjugation appear in the CPT theorem~\cite{Streater&Wightman}.

The inversion $\mathfrak{I}$ appears also as the modular conjugation operator in the Tomita-Takesaki theorem with the modular Hamiltonian represented by the dilatation $D \equiv L_{45}$~\cite{Haag}. In other words, it turns out the local algebra of 
observables $\mathcal{A}$ generated by free and real conformal quantum scalar field can be 
effectively separated into mutually independent (commuting) subalgebras $\mathcal{A}_\text{I}$ and $\mathcal{A}_\text{II}$, such that 
$\mathfrak{I}\mathcal{A}_\text{I,II}\mathfrak{I} = \mathcal{A}_\text{II,I}$ and dynamics in $\mathcal{A}_\text{I,II}$ is set by 
the dilatation $D$.\footnote{The charge conjugation acts here as the identity.} More known example of such a separation of
the local observables is given by expanding the field into the left and right Rindler modes, i.e. 
$\mathcal{A} = \mathcal{A}_\text{L}\otimes\mathcal{A}_\text{R}$ with the modular conjugation $e^{i\pi\hat{L}_{12}}\circ\mathfrak{I}$
and the modular Hamiltonian given by the boost Lorentz operator 
$B \equiv L_{03}$. From the physical point of view, 
it appears that observables composed of the half of the field degrees of freedom satisfy the Kubo-Martin-Schwinger (KMS) 
condition~\cite{Haag} in the ordinary Poincar\'{e}-invariant vacuum $|\Omega\rangle$. That is if $\hat{O}(x) \in \mathcal{A}_\text{R}$, 
then
\beqa
\langle\Omega|\hat{O}(x)|\Omega\rangle &=& \text{Tr}\big(\rho_{\text{T}_\text{U}}\hat{O}(x)\big)\,,
\eeqa
where $\rho_{\text{T}_\text{U}}$ is the density matrix corresponding to the thermal equilibrium at the Unruh (U) temperature 
$\text{T}_\text{U}$~\cite{Unruh,Sciama&Candelas&Deutsch,Takagi,Sewell,Kay}. In other words, the pure state 
$|\Omega\rangle$ appears as a thermal state when one probes it by local observables composed of the specific half of the field degrees 
of freedom.

In Section~\ref{sec:1}, I will discuss the free and real conformal field quantization in Minkowski, de Sitter (dS) and 
anti-de Sitter (AdS) spaces. It will turn out that the anti-unitary operator $\mathfrak{I}$ is also a modular conjugation 
in AdS space and up to a certain unitary operator in dS space for local subalgebras analogous to 
$\mathcal{A}_\text{I,II}$. Besides, I will consider the so-called conformal inversion operator $\mathfrak{I}_\text{C}$ 
and its representation on the solution space $\mathcal{S}$ of the conformal field equation. This operator is unitary 
and hermitian on $\mathcal{S}$~\cite{Kastrup} and it will be shown that $\mathfrak{I}_\text{C}$ annihilates 
$\mathcal{A}_\text{I,II}$. I will also obtain its eigenfunctions and identify them with the well-known field modes being eigenfunctions of $H \equiv L_{05}$.

In Section~\ref{sec:2}, I will consider various physical processes where the unitarity is violated. It occurs under the assumption that detector's state can change during those processes. In particular, if one imagines a detector that 
being inertial starts to speed up till a constant proper acceleration, then its state has to change from the ordinary 
state to the unitary inequivalent one in order to absorb Rindler particles. This implies a violation of the unitarity. However, the unitarity is preserved if one leaves the state unchanged\footnote{Up to a unitary transformation. This is also meant below when it is written
a state does not change.} and admits that the space of observables being felt by the detector
alters itself in such a way that some of them become unmeasurable or hidden. That is a local observable $\hat{O}(x)$ being used to
probe properties of a quantum state and belonging initially to $\mathcal{A}_\text{L}\otimes\mathcal{A}_\text{R}$ starts to
lie, say, in $\mathbf{\hat{1}}\otimes\mathcal{A}_\text{R}$ at some moment of time.
I will also discuss how that may perhaps resolve the black hole information paradox~\cite{Hawking,Frolov&Novikov,Brustein}.

Throughout this paper the fundamental constants are set to unity, i.e. $c = k_\text{B} = G = \hbar = 1$
by definition.

\section{Linear CFT and $\mathfrak{I}$ and $\mathfrak{I}_\text{C}$ symmetries}
\label{sec:1}

Quantization of a non-interacting field consists of several steps. One of them is to find a classical solution of the field equation. 
The solution space $\mathcal{S}$ is not positive definite with respect to the Klein-Gordon scalar product taken over Cauchy 
surface $\Sigma$. However, that space can be divided into two mutually orthogonal subspaces, i.e. 
$\mathcal{S} = {\mathcal{S}^+}\oplus\,\mathcal{S}^{-}$, such that the scalar product is positive (negative) definite on 
$\mathcal{S}^{+}$($\mathcal{S}^{-}$)~\cite{Birrell&Davies}.

The field equation depends on the spacetime metric. I will deal with only maximally symmetric spacetimes. Their line elements 
can be written down in the following form 
\beqa\label{eq:metric:mink_ds_ads}
ds^2 = a^2(x)\eta_{\mu\nu}dx^{\mu}dx^{\nu}\,,
\quad \text{where} \quad a(x) \;=\; \frac{1 + |\alpha|}{1 - \alpha x^2}\,
\eeqa
and $x^2 \equiv \eta_{\mu\nu}x^{\mu}x^{\nu}$, $\eta_{\mu\nu}$ is the Minkowski metric with mostly minus signs. The advantage 
of that metric is that it allows to consider simultaneously Minkowski, de Sitter or anti-de Sitter spaces by setting $\alpha = 0$, 
$1$ or $-1$, respectively. In case of de Sitter and anti-de Sitter spaces, one has to map $x^{\mu}$ appearing in the metric to 
$\bar{x}^{\mu} = \exp(+\frac{\pi}{2}L_{05})x^{\mu}$ and $\tilde{x}^{\mu} =\exp(+\frac{\pi}{2}L_{34})x^{\mu}$ to obtain the standard 
flat and Poincar\'e patches, respectively. The $\bar{x}^0$ coordinate corresponds to the conformal time in flat dS space with the 
conformal Killing vector $L_{05} + L_{45}$ generating translation along it. The $\tilde{x}^0$ coordinate corresponds to the time 
in the Poincar\'{e} patch of AdS space with the Killing vector $L_{05} + L_{03}$ generating $\tilde{x}^0$ translation and $\tilde{x}^3$ ranges from the horizon ($\tilde{x}^3 = \infty$) to the boundary ($\tilde{x}^3 = 0$) of AdS space.\footnote{Note that the set of Killing 
vectors in $\mathfrak{so}(2,4)$ are $\{L_{\mu 5} + L_{\mu 4}, L_{\mu\nu}\}$ for Minkowski, $\{ L_{\mu 4}, L_{\mu\nu}\}$ for de Sitter 
and $\{L_{\mu 5}, L_{\mu\nu}\}$ for anti-de Sitter spaces. The relation between $\{L_{\mu4},L_{\mu5}\}$ and $\{P_{\mu},K_{\mu}\}$ 
is given by $P_{\mu} = L_{\mu5} + L_{\mu4}$ (translation) and $K_{\mu} = L_{\mu5} - L_{\mu4}$ (conformal translation).}

In case of the free conformal field theory, one can then immediately find the normalized mode functions of the dynamical equation: 
\beqa
\Phi_{\mathbf{k}}(x) &=& \frac{1}{(2\pi)^{\frac{3}{2}}}\frac{\exp\left(-ikx\right)}{(2k)^{\frac{1}{2}}a(x)} \;\in\; \mathcal{S}^{+},
\quad \Phi_{\mathbf{k}}^{*}(x) \;\in\; \mathcal{S}^{-},
\eeqa
where star denotes the complex conjugation.
However, it is convenient to introduce spherical coordinates. 
This transformation induces a unitary map of the spaces $\mathcal{S}^{\pm}$ into themselves. That unitary operator 
$S(k,l,m|\mathbf{k})$ equals $i^lk^{-1}Y_{lm}^*(\Omega_{\mathbf{k}})\delta(k-k^{\prime})$ and provides\footnote{In the following the spherical harmonics are taken to be real.}
\beqa
\Phi_{klm}(x) &=& \int d^3\mathbf{k}^{\prime}\,S^*(k,l,m|\mathbf{k}^{\prime})\Phi_{\mathbf{k}^{\prime}}(x)
\;=\; \left(\frac{k}{\pi}\right)^{\frac{1}{2}}\frac{\exp\left(-ikx^0\right)}{a(x^0,|\mathbf{x}|)}
j_l\big(k|\mathbf{x}|\big)Y_{lm}\big(\Omega_{\mathbf{x}}\big)\,,
\eeqa
where $k \equiv |\mathbf{k}| \in \mathbb{R}^+$. The rescaled modes, i.e. $a(x)\Phi_{klm}(x)$,
can be in turn expanded via eigenfunctions of the dilatation operator $D$, i.e. $a(x)\Phi_{plm}(x)$, where $p \in \mathbf{R}$.
The unitary operator relating them is $C(p|k) = (2\pi)^{-\frac{1}{2}}k^{-ip - \frac{1}{2}}$~\cite{Tanaka&Sasaki}:
\beqa\label{eq:phi_plm_1}\nonumber
\Phi_{plm}(x) &=& \int dk\,C^*(p|k)\Phi_{klm}(x)
\\[2mm]
&=& \frac{i^l\exp\left(+\frac{\pi i}{4}\right)}{a(x^0,|\mathbf{x}|)\left(4\pi|\mathbf{x}|\right)^{\frac{1}{2}}}
\frac{\Gamma\left(1 +l + ip\right)}{\left(|\mathbf{x}|^2 - (x^0)^2\right)^{\frac{ip}{2}+\frac{1}{4}}}
P_{ip-\frac{1}{2}}^{-l-\frac{1}{2}}\left(\frac{ix^0}{\left(|\mathbf{x}|^2 - (x^0)^2\right)^{\frac{1}{2}}}\right)Y_{lm}(\Omega)\,.
\eeqa
The dilatation vector $D$ is not globally time- or spacelike, it divides space into subspaces, wherein it is
either timelike or spacelike. It means that $\Phi_{plm}(x)$ is a restriction of the ordinary modes $\Phi_{\mathbf{k}}(x)$ defined on 
whole space into those subspaces. Assuming $D$ is timelike and future-directed (the region $\textrm{I}$ in fig.~\ref{fig:1}), 
one has to set $(|\mathbf{x}|^2 - t^2)^{\frac{1}{2}} = i (t^2 - |\mathbf{x}|^2)^{\frac{1}{2}}$ in 
\eqref{eq:phi_plm_1}. 

\subsection{Representation of $\mathfrak{I}$ and $\mathfrak{I}_\text{C}$ on $\mathcal{S}$}

Among global symmetries of spacetimes under consideration, it turns out there are two which play a 
distinguished role in the quantization of the field. Specifically, the inversion $\mathfrak{I}$ and conformal inversion 
$\mathfrak{I}_\text{C}$ defined as
\beqa
\mathfrak{I}x^{\mu} &\equiv& - x^{\mu}\, \quad \textrm{and} \quad
\mathfrak{I}_\text{C}x^{\mu} \;\equiv\; - \frac{x^{\mu}}{x^2}\,.
\eeqa
These transformations provide a map of spacetime into itself which can also be understood in terms
of certain operators mapping $\mathcal{S}$ into itself.

To recover the action of $\mathfrak{I}$ on the modes $\Phi_{plm}(x)$, one has first to consider the operator 
$\mathfrak{I}\circ\mathfrak{I}_\text{C}$.  This composite operator is anti-unitary, therefore 
$\mathfrak{I}\circ\mathfrak{I}_\text{C}\Phi_{plm}(x) \in \mathcal{S}^{-}$. This means one can unitary map
$\mathfrak{I}\circ\mathfrak{I}_\text{C}\Phi_{plm}(x)$ into $\Phi_{plm}^*(x)$. Hence, one obtains
\beqa
\mathfrak{I}(p|p^{\prime}) &=& (-1)^{l+1}\delta(p+p^{\prime})\,\delta_{ll^{\prime}}\delta_{mm^{\prime}}\,\mathfrak{C}\,,
\eeqa
where $\mathfrak{C}$ is the complex conjugation operator. Hence, the operator $\mathfrak{I}$ is an anti-unitary 
on $\mathcal{S}$, i.e. $\mathfrak{I}\mathcal{S}^{\pm} =\mathcal{S}^{\mp}$.

The conformal inversion $\mathfrak{I}_\text{C}$ is represented by a unitary and 
hermitian operator on the spaces $\mathcal{S}^{\pm}$~\cite{Kastrup}. Indeed, one has
\beqa
\mathfrak{I}_\text{C}(\mathbf{k}|\mathbf{k}^{\prime}) &\equiv& 
\big(\mathfrak{I}_\text{C}\Phi_{\mathbf{k}},\Phi_{\mathbf{k}^{\prime}}\big)_{\Sigma} \;=\; -
\frac{J_0\big((2k^{\mu}k_{\mu}^{\prime})^{\frac{1}{2}}\big)}{4\pi\left(kk^{\prime}\right)^{\frac{1}{2}}}\,
\eeqa
and $\big(\mathfrak{I}_\text{C}\Phi_{\mathbf{k}},\Phi_{\mathbf{k}^{\prime}}^{*}\big)_{\Sigma} = 0$, where
$k \equiv |\mathbf{k}|$. Thus,
$\mathfrak{I}_\text{C}^{\dagger}(\mathbf{k}|\mathbf{k}^{\prime}) =\mathfrak{I}_\text{C}(\mathbf{k}|\mathbf{k}^{\prime})$
and
\beqa
\int d\mathbf{k}^{\prime\prime}\,\mathfrak{I}_\text{C}^{\dagger}(\mathbf{k}|\mathbf{k}^{\prime\prime})
\mathfrak{I}_\text{C}(\mathbf{k}^{\prime\prime}|\mathbf{k}^{\prime}) &=& \delta\big(\mathbf{k} - \mathbf{k}^{\prime}\big)\,.
\eeqa

The operator $\mathfrak{I}_\text{C}(\mathbf{k}|\mathbf{k}^{\prime})$ is an integral transform of the modes $\Phi_{\mathbf{k}}(x)$
into those lying in $\mathcal{S}^{+}$. In the following I will mostly deal with the modes 
$\Phi_{plm}(x)$ in which $\mathfrak{I}_\text{C}(\mathbf{k}|\mathbf{k}^{\prime})$ represented as
\beqa\nonumber\label{eq:conformal_inversion_plm}
\mathfrak{I}_\text{C}(p|p^{\prime}) &=&
\int dk dk^{\prime} d\mathbf{q} d\mathbf{q}^{\prime} C^*(p|k)S^*(k|\mathbf{q})
\mathfrak{I}_\text{C}(\mathbf{q},\mathbf{q}^{\prime})C(p^{\prime}|k^{\prime})S(k^{\prime}|\mathbf{q}^{\prime})
\\[1mm]
&=& - \frac{\Gamma\left(1 + l + ip\right)}{\Gamma\left(1 + l - ip\right)}\,\delta(p+p^{\prime})\,\delta_{ll^{\prime}}\delta_{mm^{\prime}}\,,
\eeqa
where $\{l,m\}$ and $\{l^{\prime},m^{\prime}\}$ parameters have been suppressed.
Since the unitary and hermitian operator $\mathfrak{I}_\text{C}$ acts on the manifolds as a transformation
of $\textrm{I}$ into $\textrm{II}$ and vice versa, the modes in $\textrm{II}$ equivalent to $\Phi_{plm}(x)$ in $\textrm{I}$
are given by $\mathfrak{I}_\text{C}\Phi_{plm}(x)$.
\begin{figure}
\includegraphics[width=4.00cm]{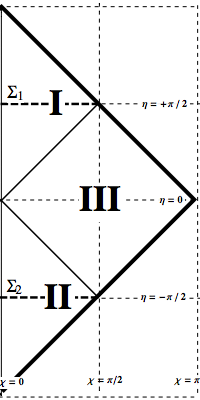}\quad\quad
\includegraphics[width=4.00cm]{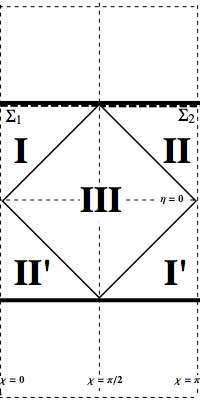}\quad\quad
\includegraphics[width=4.00cm]{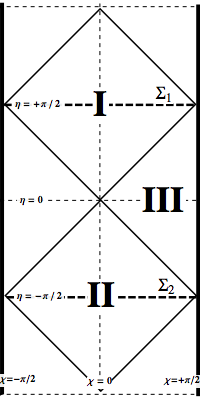}
\caption{Conformal or Penrose diagrams of Minkowski (left), de Sitter (middle) and anti-de Sitter (right) spacetimes 
embedded in the Einstein static universe. The coordinates $(\eta,\chi,\theta,\phi)$ are the so-called closed coordinates
in which the spatial section has a topology $\mathbf{S}^3$. Each point on the plots is a two-dimensional sphere 
with the coordinates $(\theta,\phi)$.}
\label{fig:1}
\end{figure}
Introducing the so-called open coordinates 
$x^0 = e^{\bar{\eta}}\cosh\bar\chi$ and $|\mathbf{x}| = e^{\bar{\eta}}\sinh\bar\chi$ (see App.~\ref{app:1}) in $\Phi_{plm}(x)$ given by
\eqref{eq:phi_plm_1} and $\mathfrak{I}_\text{C}\Phi_{plm}(x)$, one obtains
\bsubeqs
\beqa\label{eq:phi_plm}
\Phi_{plm}(x) &=& +\frac{i^le^{+\frac{\pi p}{2}}}{a(\bar{\eta},\bar\chi)}e^{-ip\bar{\eta}}
\frac{\Gamma\left(1 +l + ip\right)}{\left(4\pi\sinh\bar{\chi}\right)^{\frac{1}{2}}}
P_{ip-\frac{1}{2}}^{-l-\frac{1}{2}}\big(\cosh\bar{\chi}\big)Y_{lm}(\bar\Omega)\,,
\\[2mm]
\mathfrak{I}_\text{C}\Phi_{plm}(x) &=& -\frac{i^le^{-\frac{\pi p}{2}}}{a(\bar{\eta},\bar\chi)}e^{+ip\bar{\eta}}
\frac{\Gamma\left(1 +l + ip\right)}{\left(4\pi\sinh\bar{\chi}\right)^{\frac{1}{2}}}
P_{ip-\frac{1}{2}}^{-l-\frac{1}{2}}\big(\cosh\bar{\chi}\big)Y_{lm}(\bar\Omega)\,.
\eeqa
\esubeqs
The effective Cauchy surface $\Sigma_\text{eff} = \Sigma_1\cup\Sigma_2$ depicted in fig.~\ref{fig:1} is enough to
set well-defined Cauchy problem. The modes $\Phi_{plm}(x)$ are normalized over it in the following sense 
\beqa
(\Phi_{plm},\Phi_{p^{\prime}l^{\prime}m^{\prime}})_{\Sigma_\text{eff}} &=&
(\Phi_{plm},\Phi_{p^{\prime}l^{\prime}m^{\prime}})_{\Sigma_1} + 
(\mathfrak{I}_\text{C}\Phi_{plm},\mathfrak{I}_\text{C}\Phi_{p^{\prime}l^{\prime}m^{\prime}})_{\Sigma_2}
\;=\; \delta(p-p^{\prime})\delta_{ll^{\prime}}\delta_{mm^{\prime}}\,.
\eeqa

\subsection{Continuation of CFT from $\mathbf{R}\times\mathbf{H}^3$ to $\mathbf{R}\times\mathbf{S}^3$}

Since $\mathfrak{I}_\text{C}$ is unitary and hermitian, it possesses two eigenvalues $\pm 1$. In other words, the space 
$\mathcal{S}^{+}$ can be splitted into two orthogonal subspaces $\mathcal{S}_{+}^{+}$ and $\mathcal{S}_{-}^{+}$~\cite{Kastrup}.
The modes $\Phi_{plm}(x) \in \mathcal{S}^{+}$ can be mapped into  $\mathcal{S}_{\pm}^{+}$ by applying the projector
$\mathbf{P}_{\pm} \equiv \frac{1}{2}(\mathbf{1} \pm \mathfrak{I}_\text{C})$. 
Indeed, with the aid of \eqref{eq:conformal_inversion_plm} one gets
\beqa
\mathfrak{I}_\text{C}\big(\mathbf{P}_{\pm}\Phi_{plm}\big) &=& \pm \mathbf{P}_{\pm}\Phi_{plm}\,.
\eeqa
On the other hand, the operator $\mathfrak{I}_\text{C}$ is an integral transform and there exists $M(n|p)$, such that
$\mathfrak{I}_\text{C}\text{M}(n|p) = (-1)^{n+1}\text{M}(n|p)$, where $n$ is a nonnegative integer (see App.\ref{app:2}). 
Its normalized eigenfunctions are
\beqa\nonumber\label{eq:M}
\text{M}(n,p) &=& \frac{2^{l+1}i^{n+l}}{\Gamma\left(2+2l+n\right)^{\frac{1}{2}}}
\frac{\Gamma\left(1+l+n-ip\right)}{(2\pi)^{\frac{1}{2}}\Gamma\left(1+n\right)^{\frac{1}{2}}}
\\[1mm]
&&\times\;\frac{\Gamma\left(1+l+ip\right)}{\Gamma\left(1+l-ip\right)}\;
_2F_1\left(-n,1+l+ip,-n-l+ip;-1\right)\,.
\eeqa
Thus, one can define modes $\Phi_{nlm}(x)$ labeled by the discrete parameter $n$ ($\ge 0$) instead the continuous one 
$p$ ($\in \mathbf{R}$) as follows
\beqa
\Phi_{nlm}(x) &=& \int dp\;\text{M}^*(n|p)\Phi_{plm}(x)\,,
\eeqa
such that $\mathfrak{I}_\text{C}\Phi_{nlm}(x) = (-1)^{n+1}\Phi_{nlm}(x)$. Taking into account the action of $\mathfrak{I}_\text{C}$
on spacetime points, one can establish that $a(x)\Phi_{nlm}(x)$ are eigenfunctions of the Killing vector $H$ (see App. \ref{app:3}):
\beqa\label{eq:phi_nlm}
\Phi_{nlm}(x) &=&  \frac{i^{n+l+1}\sin^l\chi}{2^{l+1}a(\eta,\chi)}
\frac{\Gamma\left(2+2l\right)\Gamma\left(1+n\right)^{\frac{1}{2}}}{\Gamma\left(l+\frac{3}{2}\right)\Gamma\left(2+2l+n\right)^{\frac{1}{2}}}
e^{-i(n + l + 1)\eta}C_{n}^{l+1}(\cos\chi)Y_{lm}(\Omega)\,
\eeqa
expressed in the so-called closed coordinates (see App.~\ref{app:1}).

Thus, the modes $\Phi_{plm}(x)$ found in spacetime with the hyperbolic spatial section $\mathbf{H}^3$ have been related with 
those defined on the spherical one $\mathbf{S}^3$. Therefore, $\Phi_{plm}(x)$ can be continued to whole spacetime and, hence, 
to the Einstein static universe~\cite{Luescher&Mack}.

\subsection{Thermal states}

For an observer freely moving through the region $\text{I}$ and having no excess to the region $\text{II}$, the ordinary CFT
vacuum is seen as filled with the thermal bath of particles defined with respect to $D$. 
Rigorously speaking, the CFT vacuum restricted to the region $\text{I}$ is a
conformal KMS state with temperature $\text{T} = 1/2\pi a(\bar{\eta})$. 

\subparagraph*{Minkowski space}

The region $\text{I}$ in Minkowski space is identified with the expanding Milne universe. The dilatation operator $D$ is
a generator of comoving geodesics in that region. Its positive frequency eigenfunctions define the so-called
conformal Milne vacuum. The Minkowksi and Milne vacua are not unitary equivalent. An observer probing the
quantum field being in the Minkowski state has to find it as being a thermal state with temperature 
$\text{T} = e^{-\bar{\eta}}/2\pi$, where $e^{2\bar{\eta}} = x^2 \in (0,+\infty)$~\cite{Birrell&Davies}.

The region $\text{III}$ in fig.\ref{fig:1} is Rindler spacetime to be associated with the proper reference frame of a uniformly accelerated
observer in Minkowski space. Dynamics in the Rindler frame is governed by the boost Killing vector $B$ whose positive frequency
eigenfunctions define the Rindler vacuum. 
It is unitary equivalent to the Milne vacuum~\cite{Hislop&Longo}. 
Indeed, the region $\text{I}$ in the open coordinates $(\bar\eta,\bar\chi,\bar\theta,\bar\phi)$ is mapped into 
$\text{III}$ by
\beqa\label{eq:I_to_III}
e^{\bar\phi L_{12}-\bar\theta L_{13}}\,e^{\frac{\pi}{2}L_{05}-\frac{\pi}{2}L_{34}}\,e^{\bar\theta L_{13} - \bar\phi L_{12}}
\eeqa
which has a unitary implementation providing isomorphism between $\mathcal{A}_\text{I}$ and $\mathcal{A}_\text{R}$. 

This transformation can be understood as
$\bar{\eta} \rightarrow \tilde\chi - \frac{\pi i}{2}$, $\bar{\chi} \rightarrow \tilde\eta + \frac{\pi i}{2}$, $\bar\theta \rightarrow \tilde\theta$
and $\bar\phi \rightarrow \tilde\phi$, 
so that $\text{T} = e^{-\bar{\eta}}/2\pi \rightarrow e^{-\tilde\chi + \frac{\pi i}{2}}/2\pi i = e^{-\tilde\chi}/2\pi$, where
the new coordinates $(\tilde\eta,\tilde\chi,\tilde\theta,\tilde\phi)$ cover Rindler space with the line element taking the following 
form
\beqa
ds^2 &=& e^{2\tilde{\chi}}\big(d\tilde{\eta}^2 - d\tilde{\chi}^2 - \cosh^2\tilde{\eta}\,d\tilde\Omega^2\big)\,.
\eeqa
One can associate an acceleration $a^{\mu}$ to a timelike Killing vector $K^{\mu}$ as follows
\beqa\label{eq:acceleration}
a^{\mu} &=& \frac{\nabla_{K}K^{\mu}}{K^2}\,,
\eeqa
where $\nabla_{K}$ is a covariant derivative along $K^{\mu}$ and 
$K^2 = g_{\mu\nu}K^{\mu}K^{\nu}$~\cite{Narnhofer&Peter&Thirring}. 
Setting $\tilde\theta = 0$\footnote{One can always do that due to the symmetry of spacetime.} 
without loss of generality, one obtains
$|a| \equiv (- g_{\mu\nu}a^{\mu}a^{\nu})^{\frac{1}{2}} = e^{-\tilde\chi}$ for $K = B$. 
Hence, one obtains the Unruh temperature $\text{T}_\text{U} = |a|/2\pi$ measured by the accelerated 
observer~\cite{Unruh}.

\subparagraph*{De Sitter space}

The region $\text{I}$ is open dS space (spatial section is hyperbolic $\mathbf{H}^3$ like in the Milne universe). 
The comoving geodesics in $\text{I}$ are the integral curves
of the dilatation $D$. The Chernikov-Tagirov state restricted to it is a thermal state, so that a comoving observer in
open de Sitter space has to register a thermal bath of particles defined with respect to $D$ with  temperature 
$\text{T} = -\sinh\bar{\eta}/2\pi$, where now $\bar{\eta} \in (-\infty,0)$.

The region $\text{III}$ is associated with the proper reference frame of a geodesic observer ($\chi = \frac{\pi}{2}$) or a uniformly
accelerated one ($\chi \neq \frac{\pi}{2}$) in de Sitter spacetime. This is the
so-called static de Sitter space.
Dynamics inside the region $\text{III}$ is set by the Killing vector $B$. The region $\text{I}$ is mapped to the region $\text{III}$
by \eqref{eq:I_to_III}. Performing the same analytic continuation of $(\bar{\eta}, \bar\chi)$ into 
$(\tilde{\eta}, \tilde\chi)$ and setting $\tilde\theta = 0$,
one derives the Narnhofer-Peter-Thirring (NPT) temperature 
$\text{T}_\text{NPT} = \cosh{\tilde{\chi}}/2\pi = (|a| + 1)^{\frac{1}{2}}/2\pi$~\cite{Narnhofer&Peter&Thirring}. It diverges on the
horizons and reduces to the Gibbons-Hawking (GH) temperature $\text{T}_\text{GH} = 1/2\pi$ for the geodesic
observer~\cite{Gibbons&Hawking}.

\subparagraph*{Anti-de Sitter space}

This case is mostly a repetition of the Minkwoski and de Sitter ones, wherein one takes the AdS state as a
physical vacuum~\cite{Avis&Isham&Storey}. 
The value of temperature merely changes due to the difference between scale factors of the spaces \eqref{eq:metric:mink_ds_ads} 
and equals $\cosh\bar{\eta}/2\pi$ in the region $\text{I}$ of AdS space, where $\bar{\eta} \in (-\infty,+\infty)$~\cite{Emelyanov}. 

The region $\text{III}$ is filled by the integral curves of $B$ associated with the observer moving with a constant 
acceleration. Performing the analytic continuation of the coordinates $(\bar{\eta},\bar{\chi})$ as above, one obtains
the Deser-Levin (DL) temperature $\text{T}_\text{DL} = \sinh\tilde\chi/2\pi = (|a|^2 -1)^{\frac{1}{2}}/2\pi$ registered by 
the uniformly accelerated detector~\cite{Deser&Levin}.
The AdS horizons and boundary are at $\tilde\chi = +\infty$ and $\tilde\chi = 0$, respectively. 
Thus, temperature $\text{T}_\text{DL}$ is divergent on the horizons and vanishes on the boundary.

One can now straightforwardly generalize the results found in Minkowski space in~\cite{Hislop&Longo} to 
de Sitter and anti-de Sitter spaces for the free conformal field theory. Specifically, the conformal vacuum defined in the
region $\text{I}$ of the dS (AdS) hyperboloid is unitary equivalent to the vacuum state defined in the region $\text{III}$
of dS (AdS) space.

\subsection{Operators $\mathfrak{I}$ and $\mathfrak{I}_\text{C}$ and inequivalent quantization}

The modes defining the conformal vacua in the regions $\text{I}$ and $\text{II}$ and normalized over $\Sigma_1$ 
and $\Sigma_2$ in Minkowski spacetime are
\bsubeqs\label{eq:mink:modes_I&modes_II}
\beqa
\Phi_{\omega lm}^\text{I}(x) &=& \frac{\exp\left(+\frac{\pi\omega}{2}\right)}{\left(2\sinh\pi\omega\right)^{\frac{1}{2}}}
\big(\Phi_{+\omega lm}(x) - (-1)^le^{-\pi\omega}\Phi_{-\omega lm}^*(x)\big)\,,
\\[0.5mm]
\Phi_{\omega lm}^\text{II}(x) &=& \frac{\exp\left(+\frac{\pi\omega}{2}\right)}{\left(2\sinh\pi\omega\right)^{\frac{1}{2}}}
\big(\Phi_{-\omega lm}^*(x) - (-1)^le^{-\pi\omega}\Phi_{+\omega lm}(x)\big)\,,
\eeqa
\esubeqs
where $\omega \equiv |p|$. According to the Tomita-Takesaki theorem specialized to the present case~\cite{Haag}, 
$\mathfrak{I}$ is the modular conjugation operator. In particular, it means that $\mathfrak{I}$ maps 
$\Phi_{\omega lm}^\text{I}(x)$ into $\Phi_{\omega lm}^\text{II}(x)$ and vice versa: 
\beqa
\mathfrak{I}\Phi_{\omega lm}^\text{I}(x) &=& \Phi_{\omega lm}^\text{II}(x)\, 
\quad \textrm{and} \quad 
\mathfrak{I}_\text{C}\Phi_{\omega lm}^\text{II}(x) \;=\; \Phi_{\omega lm}^\text{I}(x)\,.
\eeqa
Since $\Phi_{\omega lm}^\text{I}(x)$ and $\Phi_{\omega lm}^\text{II}(x)$ have zero supports in regions $\text{II}$ and $\text{I}$, respectively,  they must be, however, annihilated by the operator $\mathfrak{I}_\text{C}$. Indeed, using \eqref{eq:conformal_inversion_plm}, one obtains
\beqa
\mathfrak{I}_\text{C}\Phi_{\omega lm}^\text{I}(x) &=& 0\, 
\quad \textrm{and} \quad 
\mathfrak{I}_\text{C}\Phi_{\omega lm}^\text{II}(x) \;=\; 0\,.
\eeqa

These results can be immediately generalized to dS and AdS spaces. The action of the operators $\mathfrak{I}$
and $\mathfrak{I}_\text{C}$
do not change on the space $\mathcal{S}$. However, its action on the spacetime points slightly differs in dS and AdS 
spaces from that in Minkowski one. In terms of the closed coordinates $(\eta,\chi,\theta,\phi)$ (see App.~\ref{app:1}) covering 
the whole spaces under consideration, one finds
\beqa
\mathfrak{I}(\eta,\chi,\theta,\phi) &=&
\left\{
\begin{array}{cl}
(-\eta,\chi,\pi-\theta,\pi+\phi) & \text{-- de Sitter space}\,,
\\[1mm]
(2\pi k - \eta,\chi,\pi-\theta,\pi+\phi) & \text{-- anti-de Sitter space}\,,
\end{array}
\right.
\eeqa
and
\beqa
\mathfrak{I}_\text{C}(\eta,\chi,\theta,\phi) &=&
\left\{
\begin{array}{cl}
(\eta,\pi - \chi,\theta,\phi) & \text{-- de Sitter space}\,,
\\[1mm]
(\eta-\pi + 2\pi k,\chi,\pi - \theta,\pi + \phi)  & \text{-- anti-de Sitter space}\,,
\end{array}
\right.
\eeqa
where $k \in \mathbb{Z}$.

\subparagraph*{De Sitter space}

The operator $\mathfrak{I}$ maps $\text{I}$ into $\text{II}^{\prime}$. However, the region $\text{II}^{\prime}$ can in turn
be mapped to the region $\text{II}$ by $e^{\pi H/2}$. This mapping has a unitary implementation on $\mathcal{S}$. 
In other words, the modes $e^{-i\pi \hat{H}/2}\circ\mathfrak{I}_\text{C}\Phi_{plm}(x)$ have a nonzero 
support in $\text{II}^{\prime}$ and define the CFT vacuum in that region like $\Phi_{plm}(x)$ and 
$\mathfrak{I}_\text{C}\Phi_{plm}(x)$ do that in $\text{I}$ and $\text{II}$, respectively.

One can define modes $\Phi_{\omega lm}^\text{I}(x)$ and $\Phi_{\omega lm}^\text{II}(x)$ being
analogous to \eqref{eq:mink:modes_I&modes_II} in Minkowski space and
defining the conformal vacua in $\text{I}$ and $\text{II}$, respectively. The modular conjugation mapping 
$\Phi_{\omega lm}^\text{I,II}(x)$ into $\Phi_{\omega lm}^\text{II,I}(x)$ is given by $e^{i\pi \hat{H}/2}\circ\mathfrak{I}$.
The operator $\mathfrak{I}_\text{C}$ annihilates both of them.

\subparagraph*{Anti-de Sitter space} The only difference between the operators $\mathfrak{I}$ and $\mathfrak{I}_\text{C}$ 
in anti-de Sitter and Minkowski spaces is that one has infinitely many wedge regions equivalent to $\text{I}$ and $\text{II}$ in AdS
space.
The reason lies in that the topology of the AdS hyperboloid is $\mathbf{S}\times\mathbf{R}^3$, where the time coordinate $\eta$
runs over the circle $\mathbf{S}$. This leads to the existence of the closed timelike curves. One
usually unwraps $\mathbf{S}$ and deals with its universal covering $\mathbf{R}$ to avoid casual paradoxes, i.e.
$\eta \in (-\pi,+\pi) \rightarrow \eta \in (-\infty,+\infty)$. Therefore, one has infinitely many wedge regions equivalent to
either $\text{I}$ or $\text{II}$ from the CFT point of view.

\section{Non-unitarity and hidden observables}
\label{sec:2}

\subsection{Violation of unitarity}

For a quantization of the field $\Phi(x)$ and the concepts of vacuum and particle the symmetries of spacetime play a crucial role. Nevertheless, one can imagine an observer who moves through spacetime not along a (conformal) Killing vector $K$, but along 
a certain vector field $V$. This vector $V$ sets dynamics in observer's reference frame. Since, in general, it is not one of $K$ 
attributed to spacetime, there is no conserved quantity associated with it. However, if there are time intervals during which $V$ 
is roughly equal to $K$, there has to appear a conserved quantity (Hamiltonian) approximately equaling to that associated with 
$K$. During these time intervals, the field excitations are naturally defined by expanding the field through the positive and negative frequency modes of $K$, i.e. $\Phi_{\omega}(x)$ and $\Phi_{\omega}^*(x)$, s.t.
\beqa
K\big(a(x)\Phi_{\omega}(x)\big) \;=\; -i\omega\big(a(x)\Phi_{\omega}(x)\big)\,,
\eeqa
where $\omega \in \mathbf{R}^{+}$ is interpreted as being the energy of the excitation and the rest of indices counting the field 
degrees of freedom have been suppressed. 

Specifically, one may consider a detector moving along
\beqa\label{eq:noninertial_v}
V &\rightarrow&
\left\{
\begin{array}{ll}
B\,, & x^0 \;\rightarrow\; +\infty\,, \\
P_0\,, & x^0 \;\rightarrow\; -\infty\,,
\end{array}
\right.
\eeqa
where $B = L_{03} = x^0\partial_{x^3} + x^3\partial_{x^0}$ and $P_0 = L_{05} + L_{04} = \partial_{x^0}$.
At past-time infinity, the detector has presumably to register no particles, i.e. its state is
the ordinary Minkowski vacuum. At future-time infinity, the detector has to register the thermal bath with temperature
$\text{T}_\text{U}$. The operators $\hat{P}_0$ and $\hat{B}$ can be mapped into each other, but one has to use
a non-unitary operator for that, namely
\beqa
-ie^{-i\pi\hat{L}_{34}/2}e^{-\pi\hat{D}/2}{:}&&
\hat{P}_0 \;\rightarrow\; \hat{B}\,.
\eeqa
This case has to be distinguished from that when $V \rightarrow D$ at $x^0 \rightarrow -\infty$, because $\hat{D}$ can 
be unitary mapped to $\hat{B}$. This means that if one sets detector's state at past-time infinity to be the conformal 
Milne one, then the detector along its movement would measure temperature gradually increasing from $0$ to 
$\text{T}_\text{U}$ with no violation of unitarity.

It is worth mentioning another example illustrating what has been meant. One may imagine a universe evolving
from Minkwoski space to de Sitter space. This is realized by taking, for instance, the flat FRW metric with the scale factor 
$a(\eta) = 1- 1/\eta$, where $\eta \in (-\infty, 0)$ is the conformal time and the de Sitter curvature has been set to 
unity.
One can set that at $|\eta| \gg 1$, a comoving detector is in the Minkowski state becoming the Chernikov-Tagirov
one at $\eta$ approaching $0$ (both are the ordinary CFT vacuum here). However, its reference frame at $|\eta| \ll 1$ 
is restricted to de Sitter space in the static coordinates. That is detector's state would change non-unitary if it can
absorb particles defined with respect to the static dS vacuum (positive energy excitations in static dS space defined with
respect to $L_{04}$). The unitarity is not violated if detector's state at past-time infinity is not the Minkowski state, 
but the conformal Milne one. Nevertheless, the effect is the same in that sense the detector would measure temperature 
increasing from $0$ to $\text{T}_\text{GH}$:
\beqa
\text{T} &=& \frac{1}{2\pi\big(1 - \eta\big)} \; \rightarrow\;
\left\{
\begin{array}{ll}
0\,, & \eta \;\rightarrow -\infty\,,
\\[1.5mm]
\frac{1}{2\pi}\,, & \eta \;\rightarrow\; -0\,,
\end{array}
\right.
\eeqa
assuming the detector is located at the spatial origin. Note that in this case $V \rightarrow D$ at past-time infinity and
$V \rightarrow L_{04}$ at future-time infinity. The dilatation $\hat{D}$ and $\hat{L}_{04}$ can be mapped into each other 
by the unitary operator generated by $\hat{H}$.

It is important to distinguish these cases from particle production in a universe evolving in time $t$.
To have particle production, one has explicitly to break 
the conformal symmetry by adding, for example, a mass term. This results in the time-dependence of the frequency $\omega$. 
Assuming that $\omega \rightarrow \omega_\text{in} = \text{const}$ at past-time infinity and 
$\omega \rightarrow \omega_\text{out} = \text{const} \neq \omega_\text{in}$ at future-time infinity, the normalized modes 
$\Phi_1(x)$ solving the field equation and approaching $\Phi_\text{in}(x) \propto e^{-i\omega_\text{in}t}$ at 
$t \rightarrow -\infty$ become a linear combination 
of $\Phi_\text{out}(x)$ and $\Phi_\text{out}^*(x)$ at $t \rightarrow +\infty$, where 
$\Phi_\text{out}(x) \propto e^{-i\omega_\text{out}t}$ is obtained from another 
solution $\Phi_2(x)$ of the field equation at that limit. The map between $\{\Phi_1(x),\Phi_1^*(x)\}$ and 
$\{\Phi_2(x),\Phi_2^*(x)\}$ is perfectly unitary and known as the squeezed transformation.
Note that this process is similar to the Schwinger effect, i.e. the $\text{e}^+\text{e}^-$ pair production by the static electric field
(see, for example,~\cite{Anderson&Mottola} and references therein).

To sum it up, if detector's state can vary between the Minkowski and the thermal states, then one encounters a violation
of the unitarity. 

\subsection{Hidden field degrees of freedom}

To preserve the unitarity one may admit that detector's state can be identified with the thermal state (non-CFT vacuum). 
I will discuss below another possibility, i.e. whether its state identified with CFT one could stay 
unchanged with still measuring well-known thermal effects.

\subparagraph*{Inertial detector}
Consider an inertial detector $\mathbf{D}_1$ which internal dynamics is set to be governed by
\beqa
V_1 &=&
\left\{
\begin{array}{lcll}
P_0\,, & |x^0| &\geq& 1\,,
\\[1mm]
R\,, & |x^0|&<& 1\,,
\end{array}
\right.
\eeqa
where $R \equiv L_{04}$. The detector $\mathbf{D}_1$ is not excited at $|x^0| \geq 1$. However, its internal dynamics considerably changes at $x^0 = \pm1$, such that the concept of the physical field excitation becomes different during $x^0 \in (-1,+1)$ from the ordinary one. One can show that its proper reference frame coincides up to a conformal factor with the open FRW universe at $x^0 \in (-1,+1)$. Hence, one immediately obtains that $\mathbf{D}_1$ has to be thermally excited with temperature
\beqa
\textrm{T}_{\mathbf{D}_1} &=& \frac{1/\pi}{1-(x^0)^2}\,,
\eeqa
where it has been assumed the detector is located at the spatial origin~\cite{Hislop&Longo,Haag}. 

\begin{figure}
\includegraphics[width=4.00cm]{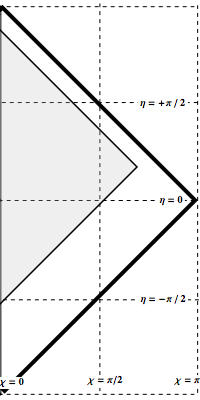}\quad\quad
\includegraphics[width=4.00cm]{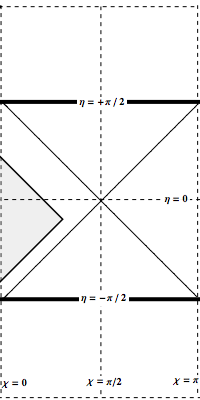}\quad\quad
\includegraphics[width=4.00cm]{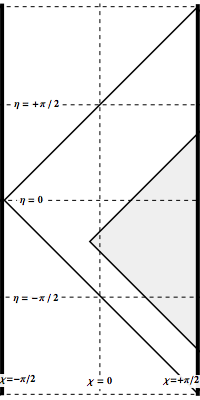}
\caption{Left: the shaded region in Minkwoski space is filled by the integral curves of $R^{\prime}$. This region can be
transformed in the contracting ($\text{II}$ depicted in fig.\ref{fig:1}) or expanding ($\text{I}$ depicted in fig.\ref{fig:1}) Milne universes as
well as it can cover almost whole space at $\alpha \rightarrow -\infty$. Middle: the shaded region in dS space is filled also by
the curves of $R^{\prime}$ with $\alpha \geq 0$. It can be mapped to open de Sitter by setting $\beta = \mp\frac{\pi}{2}$ 
($\text{I}$ or $\text{II}^{\prime}$ depicted in fig.\ref{fig:1}, respectively). Right: the shaded region in the Poincar\'{e} patch of AdS space.
It is filled by the integral curves of the Killing vector $B^{\,\prime}$ which reduces in particular to the region $\text{III}$ depicted in
fig.\ref{fig:1} if $\alpha = \beta = 0$.}
\label{fig:2}
\end{figure}

The time interval $(-1,+1)$ can be stretched to almost whole Minkowski space. Indeed, the hermitian operator 
$\hat{R}$ is unitary transformed to the linear combination of $\hat{D}$, $\hat{H}$ and itself as 
follows
\beqa\label{eq:r_as_d_h_r}
\hat{R}^{\,\prime} &=& \big(e^{i\alpha\hat{D}}e^{i\beta\hat{H}}\big)\hat{R}\big(e^{i\alpha\hat{D}}e^{i\beta\hat{H}}\big)^{\dagger}
\;=\; \big(\hat{R}\cosh\alpha - \hat{H}\sinh\alpha\big)\cos\beta - \hat{D}\sin\beta\,,
\eeqa 
where $\alpha \in \mathbf{R}$ and $\beta \in [-\frac{\pi}{2},+\frac{\pi}{2}]$.
This means, in particular, that one can map $(-1,+1)$ to the expanding or contracting Milne universe 
(see fig.\ref{fig:2}) by setting $\beta = -\pi/2$ or $\beta = +\pi/2$, respectively. 
This combination of operators approaches $\hat{P_0}$ in the limit $\alpha \rightarrow -\infty$. In other words, the integral 
curves of $R^{\,\prime}$ fills out almost the whole space at that limit. If internal dynamics of the detector $\mathbf{D}_1$ is 
now set by $R^{\,\prime}$ given in \eqref{eq:r_as_d_h_r}
during $x^0 \in (x_\text{i}^0,x_\text{f}^0) = (-e^{-\alpha}(\sec\beta + \tan\beta),+e^{-\alpha}(\sec\beta - \tan\beta))$, then it will 
be thermally excited with the following temperature
\beqa\label{eq:temperature_of_r_as_d_h_r}
\text{T}_{\mathbf{D}_1} &=& \frac{1/\pi}{\left(e^{-\alpha} - e^{+\alpha}(x^0)^2\right)\cos\beta - 2x^0\sin\beta}\,.
\eeqa
At the limit $\alpha \rightarrow -\infty$ and
assuming $\beta \neq \pm \pi/2$, $\text{T}_{\mathbf{D}_1}$ vanishes at fixed $x^0$ strongly inside its range. With no respect 
of the value of $\alpha$, this temperature is divergent on the boundary values of $x^0$. 

If one describes this process as a change of detector's state, then one has a breakdown of the unitarity. 
Indeed, its state is supposed to be the Minkowski vacuum at $x^0 \notin (x_\text{i}^0,x_\text{f}^0)$. Inside that time 
interval, it has to be changed to the nonstandard vacuum with respect to which the nonstandard field excitations have been 
defined above. Thus, the pure state becomes thermalized. However, one can still have the thermal effect without changing 
detector's state and defining a new vacuum. To illustrate this, I will consider below several examples.

The frequency spectrum of the quantum fluctuations being measured by a detector probing all field degrees of freedom 
can be found by exploiting the formula
\beqa
\mathcal{P}(\omega) &=& \int\limits_{-\infty}^{+\infty}d\tau\,e^{-i\omega x^0(\tau)}
\langle\Omega|\hat{\Phi}(x^0(\tau))\hat{\Phi}(0)|\Omega\rangle\,,
\eeqa
where $\tau$ is the proper time and $\omega = k$ frequency defined with respect to 
$x^0$~\cite{Sciama&Candelas&Deutsch}.\footnote{Note that the Wightman two-point function
$\langle\Omega|\hat{\Phi}(x)\hat{\Phi}(y)|\Omega\rangle$ 
equals $-\frac{(4\pi^2a(x)a(y))^{-1}}{(x^0 - y^0 - i\varepsilon)^2-(\mathbf{x} - \mathbf{y})^2}$.} 
A geodesic detector $\mathbf{D}_1$ being sensitive to all field degrees and moving along $P_0$ is not excited by the 
fluctuations: $\mathcal{P}_{\mathbf{D}_1}(\omega) = - \frac{\omega}{2\pi}\,\theta(-\omega) = 0$. Suppose a detector $\mathbf{D}_2$ 
is not oblivious only to those field degrees to be defined with respect to $R^{\,\prime}$ inside $(x_\text{i}^0,x_\text{f}^0)$ and 
moves along $P_0$ as well. Its power spectrum is given by
\beqa
\mathcal{P}_{\mathbf{D}_2}(\omega) &=& \int\limits_{-\infty}^{+\infty}d\tau\,e^{-i\omega \bar\eta^{\prime}(\tau)}
\langle\Omega|\hat{\Phi}(\bar\eta^{\prime}(\tau))\hat{\Phi}(0)|\Omega\rangle\,,
\eeqa
where $\omega$ now is the frequency defined with respect to $R^{\,\prime} = \partial/\partial{\bar\eta^{\prime}}$.
It is not vanishing only during $x^0 \in (x_\text{i}^0,x_\text{f}^0)$. Evaluating it, one obtains
\beqa
\mathcal{P}_{\mathbf{D}_2}(\omega) &=& 4e^{-\alpha}\cos^2(\beta/2)\,\frac{\omega}{e^{2\pi\omega} - 1}\,.
\eeqa
Expressing $\omega$ through the physical frequency $\omega_\text{ph}$, one finds temperature
\eqref{eq:temperature_of_r_as_d_h_r} ascribed to the spectrum. Thus, such a detector would indicate 
a presence of the thermal bath, although the state has not been changed.

\subparagraph*{Non-inertial detector}
The quantum field $\hat{\Phi}(x)$ can be expanded through the ordinary plane waves being the eigenfunctions of $P_0$ or, equivalently, through the eigenfunctions of $B$ as follows
\beqa
\hat{\Phi}_{\mathbf{k}}(x) &=& \int dp\big(\hat{\Phi}_{p\mathbf{k}}(x) +\hat{\Phi}_{p\mathbf{k}}^{\dagger}(x)\big)\,,
\quad B\hat{\Phi}_p(x) \;=\; -ip\hat{\Phi}_p(x)\,,
\eeqa
where $p \in \mathbf{R}$, $\mathbf{k} = (k_1,k_2)$ and $\hat{\Phi}_{p\mathbf{k}}(x) = \Phi_{p\mathbf{k}}(x)\hat{a}_p$. 
One can then introduce $\omega = |p|$ and
$\Phi_{\omega\mathbf{k}}^{\text{R}}(x)$ and $\Phi_{\omega\mathbf{k}}^{\text{L}}(x)$ to be positive frequency modes
with respect to $B$, such those
\bsubeqs
\beqa
\Phi_{\omega\mathbf{k}}^\text{R}(x) &=&
\alpha_{\omega}\Phi_{+\omega+\mathbf{k}}(x) + \beta_{\omega}\Phi_{-\omega-\mathbf{k}}^*(x)\,,
\\[1mm]
\Phi_{\omega\mathbf{k}}^\text{L}(x) &=&
\alpha_{\omega}\Phi_{-\omega+\mathbf{k}}(x) +\beta_{\omega}\Phi_{+\omega-\mathbf{k}}^*(x)\,,
\eeqa
\esubeqs
where the Bogolyubov coefficients $\alpha_{\omega} = (1-e^{-2\pi\omega})^{\frac{1}{2}}$ and 
$\beta_{\omega} = -e^{-\pi\omega}\alpha_{\omega}$.
These modes $\Phi_{\omega\mathbf{k}}^\text{R}(x)$ and $\Phi_{\omega\mathbf{k}}^\text{L}(x)$ vanish 
in the left and right Rindler wedges, respectively. The field expanded through them takes the following form
\beqa
\hat{\Phi}_{\mathbf{k}}(x) &=& \int d\omega\big(\hat{\Phi}_{\omega\mathbf{k}}^\text{R}(x) 
+ \hat{\Phi}_{\omega\mathbf{k}}^{\text{R}\dagger}(x) + 
\hat{\Phi}_{\omega\mathbf{k}}^\text{L}(x) + \hat{\Phi}_{\omega\mathbf{k}}^{\text{L}\dagger}(x)\big)\,.
\eeqa

Consider a detector $\mathbf{D}_3$ moving along $V$ as in \eqref{eq:noninertial_v}, 
such that it intersects the line $x^0 = -x^3$ at a certain time moment $x^0$. 
According to Davies-Unruh effect, one expects that the modes the detector $\mathbf{D}_3$ can in principle 
probe are given by
\beqa
\Phi_{p\mathbf{k}}^{\text{D}_3}(x) &=& \delta_{+\omega,p}\Phi_{+\omega\mathbf{k}}^\text{Int}(x) + 
\delta_{-\omega,p}\Phi_{-\omega\mathbf{k}}^\text{Int}(x)\,,
\eeqa
where delta is the Kronecker symbol and  the interpolating modes have to satisfy the following conditions:
\beqa
\Phi_{+\omega\mathbf{k}}^\text{Int}(x) &=&
\left\{
\begin{array}{ll}
\alpha_{\omega}\Phi_{\omega+\mathbf{k}}^\text{R}(x) -
\beta_{\omega}\Phi_{\omega-\mathbf{k}}^{\text{L}*}(x)\,, & x^0 \;\rightarrow \; -\infty\,,
\\[1mm]
\Phi_{\omega+\mathbf{k}}^\text{R}(x)\,, & x^0 \;\rightarrow \; +\infty\,,
\end{array}
\right.
\eeqa
and
\beqa
\Phi_{-\omega\mathbf{k}}^\text{Int}(x) &=&
\left\{
\begin{array}{ll}
\alpha_{\omega}\Phi_{\omega+\mathbf{k}}^\text{L}(x) -
\beta_{\omega}\Phi_{\omega-\mathbf{k}}^{\text{R}*}(x)\,, & x^0 \;\rightarrow \; -\infty\,,
\\[1mm]
0\,, & x^0 \;\rightarrow \; +\infty\,.
\end{array}
\right.
\eeqa
The qualitative picture of the process can be described as follows. Initially, at $x^0 \rightarrow -\infty$, 
the physical field
excitations are the ordinary Minkowski particles, i.e. the detector can feel all field degrees of
freedom. However, after having intersected the line $x^0 = -x^3$, the half of them become unavailable
to the detector. At future-time infinity, $x^0 \rightarrow +\infty$, the rest of the degrees can be felt as a new kind of
the field excitations, namely the Rindler particles.
The definition of a particle implies an introduction of a no-particle state, i.e. vacuum. If detector's
state varies from the Minkowski vacuum to the Rindler state, then one has a violation of the unitarity.

One may assume that detector's state does not change. However, there are no half of the field 
degrees of freedom corresponding to the left Rindler modes in the right Rindler wedge. The frequency spectrum 
of the detector $\mathbf{D}_3$ measuring the quantum fluctuations are
\beqa\label{eq:pd2}
\mathcal{P}_{\mathbf{D}_3}(\omega) &=&
\left\{
\begin{array}{llll}
-\frac{\omega}{2\pi}\,\theta(-\omega)\,, & x^0 &\rightarrow& -\infty\,,
\\[2mm]
\frac{2\omega}{\pi |a|}(e^{2\pi\omega}-1)^{-1}\,, & x^0 &\rightarrow& +\infty\,,
\end{array}
\right.
\eeqa
where $|a|$ is the acceleration at $x^0 \rightarrow + \infty$ and it has been assumed that the periods of both inertial 
and uniformly accelerated motions are sufficiently large to have equality in \eqref{eq:pd2}. In terms of the physical 
frequency $\omega_\text{ph}$, one obtains temperature $\text{T}_\text{U}$ ascribed to the spectrum at future-time 
infinity. Although this can be further interpreted as there are some sort of the new (Rindler) particles with the energy 
$\omega$ which the detector accumulates till reaching the equilibrium stationary state, one has to refrain from such 
interpretation, otherwise the unitarity is violated.

Suppose one can construct a detector $\mathbf{D}_4$ that probes the right Rindler modes 
$\Phi_{\omega lm}^\text{R}(x)$ and is not sensitive to the left ones $\Phi_{\omega lm}^\text{L}(x)$.
One may now ask a question: What would it measure assuming it moves along $P_0$ at $\mathbf{x} = 0$?
The frequency spectrum of the quantum fluctuation measured by the detector $\mathbf{D}_4$ is
\beqa
\mathcal{P}_{\mathbf{D}_4}(\omega) &=& 
\int\limits_{-\infty}^{+\infty}d\tau\,e^{-i\omega \bar\eta(\tau)}
\langle\Omega|\hat{\Phi}(\bar\eta(\tau))\hat{\Phi}(0)|\Omega\rangle \;=\;
\frac{4\omega}{e^{2\pi\omega} - 1}\,,
\eeqa
where it has been taken into account that $\Phi_{\omega lm}^\text{R}(x)$ can be analytically 
continued from the region $\text{III}$ to the regions $\text{I}$ and $\text{II}$. Temperature 
ascribed to the spectrum is equal to
\beqa
\text{T}_{\mathbf{D}_4} &=& \frac{1}{2\pi |x^0|}\,.
\eeqa
This temperature is time-dependent and diverges at $x^0 = 0$. Physically, the detector cannot 
measure energies higher than a threshold one $E_\text{c}$. Therefore, the maximal temperature would be equal to 
$E_\text{c}/2\pi$. Thus, the detector would show a thermal distribution that could be explained without absorption of 
the Rindler particles, i.e. without changing detector's state. The detector responses non-trivially to the quantum 
fluctuation being always present in the vacuum $|\Omega\rangle$~\cite{Sciama&Candelas&Deutsch}.

\subparagraph*{De Sitter space}

In comparison with Minkowski space, there is a horizon in de Sitter space. A geodesic detector
is always oblivious to the half of the field degrees of freedom. The frequency spectrum of the detector is 
always thermal with the Gibbons-Hawking temperature: 
\beqa
\mathcal{P}(\omega) &=& \int\limits_{-\infty}^{+\infty} dt\,
e^{-i\omega t} \langle\Omega|\hat{\Phi}(t)\hat{\Phi}(0)|\Omega\rangle
\; = \; \frac{1}{2\pi}\frac{\omega}{e^{2\pi\omega}-1}\,,
\eeqa
where it has been taken into account that $x^0 = \tanh t$, where $t$ is the physical time~\cite{Sciama&Candelas&Deutsch}.

The observer could define the static vacuum being inequivalent to the Chernikov-Tagirov or Bunch-Davies 
states~\cite{Birrell&Davies}. This is perfectly fine, if one presumes, however, that the space is eternal.
Imagine a universe which is like Minkowski space at past- and future-time infinities and resembles
de Sitter space in-between. As above, there are two cases. In the first case, the state of a comoving
detector changes and then one has a violation of the unitarity. In the second case, one does not change
the state. This detector could still measure the Gibbons-Hawking temperature during
the de Sitter space phase for the frequencies $\omega \gtrsim H_\text{dS}$ during $H_\text{dS}\Delta t \gg 1$,
where $\Delta t$ is a physical time interval during which the universe resembles de Sitter space with the 
curvature $H_\text{dS}$. Indeed, the frequency spectrum is
\beqa\nonumber
\mathcal{P}(\omega) &=& \int\limits_{-\Delta t/2}^{+\Delta t/2} dt\,
e^{-i\omega t} \langle\Omega|\hat{\Phi}(t)\hat{\Phi}(0)|\Omega\rangle
\\[1mm]
&=& \frac{1}{2\pi}\frac{\omega}{e^{2\pi\omega/H_\text{dS}}-1} + \sum_{n = 1}^{+\infty}
\frac{H_\text{dS}n^2\cos\left(\omega\Delta t/2\right) - n\omega\sin\left(\omega\Delta t/2\right)}
{2\pi^2\left(n^2 + (\omega/H_\text{dS})^2\right)\exp\left(nH_\text{dS}\Delta t/2\right)}\,,
\eeqa
where the second term is negligible with respect to the first term under those conditions.

Returning to eternal de Sitter space, one may consider a geodesic detector $\mathbf{D}_5$ that is sensitive to the
modes defined with respect to $R^{\,\prime}$ during $(x_\text{i}^0,x_\text{f}^0)$ and with $\alpha \geq 0$.
Temperature would be
\beqa
\text{T}_{\mathbf{D}_5} &=& \frac{(1-(x^0)^2)/2\pi}{\left(e^{-\alpha} - e^{+\alpha}(x^0)^2\right)\cos\beta - 2x^0\sin\beta}\,.
\eeqa
Note that $\text{T}_{\mathbf{D}_5}$ reduces to the Gibbons-Hawking temperature $\text{T}_\text{GH}$ 
when $\beta = 0$ and $\alpha = 0$. This temperature $\text{T}_{\mathbf{D}_5}$ reduces to 
$-\sinh\bar{\eta}/2\pi$ when $\beta = -\frac{\pi}{2}$ and $\alpha = 0$,
where $x^0 = e^{\bar{\eta}}$ at $\mathbf{x} = 0$ and $\bar{\eta} \in (-\infty,0)$.

\subparagraph*{Anti-de Sitter space}

An observer moving along $H$ at $\chi = 0$ in AdS space is geodesic. Analogously, one may consider
a detector $\mathbf{D}_6$ that probes merely the half of the field degrees of freedom defined with respect to
$R^{\,\prime}$ and measures the frequency spectrum of the quantum fluctuations. Then it would register 
the thermal distribution with temperature
\beqa
\text{T}_{\mathbf{D}_6} &=& \frac{(1+(x^0)^2)/2\pi}{\left(e^{-\alpha} - e^{+\alpha}(x^0)^2\right)\cos\beta - 2x^0\sin\beta}\,.
\eeqa
If one sets $\beta = -\pi/2$, then this domain of AdS space is the region $\text{I}$ in fig.~\ref{fig:1}, i.e.
the so-called open AdS space.
The temperature $\text{T}_{\mathbf{D}_6}$ reduces to $\cosh\bar\eta/2\pi$, where 
$x^0 = e^{\bar\eta}$ at $\chi = 0$.

One can also consider a detector moving along $H + B = \partial/\partial\tilde{x}^0$, i.e. the time-translation
Killing vector in the Poincar\'{e} patch. If this detector is not oblivious to the positive and negative frequency 
modes defined with respect to
\beqa\label{eq:b_as_r_h_b}
\hat{B}^{\,\prime} &=&  \big(e^{i\alpha\hat{L}_{35}}e^{i\beta\hat{H}}\big)\hat{B}
\big(e^{i\alpha\hat{L}_{35}}e^{i\beta\hat{H}}\big)^{\dagger}
\;=\; \big(\hat{B}\cosh\alpha + \hat{H}\sinh\alpha\big)\cos\beta + \hat{L}_{35}\sin\beta\,.
\eeqa
then temperature of the frequency spectrum is not zero at $\tilde{x}^3 \neq 0$.
This generalizes the result found in~\cite{Deser&Levin}.

\subsection{Discussion}

I have considered above various examples when the unitarity is violated by allowing a detector to absorb 
particles which are not defined with respect to the ordinary state $|\Omega\rangle$. The reason is that 
detector's state belonging to the ordinary Fock space must non-unitary change to register particles 
defined with respect to the thermal state. Assuming that detector's state can be prepared to be
equivalent to $|\Omega\rangle$, one can still measure the well-known thermal effects which are due
to the quantum fluctuations of the field.

The vacuum activity is also probed by the vacuum expectation value of the energy-momentum
tensor of the field $\langle\Omega |\hat{T}_{\nu}^{\mu}(x)|\Omega\rangle$. It is divergent as a result of the 
distributional nature of the quantum field. After appropriate renormalization, it becomes
\beqa\label{eq:vev_emt_ren}
\langle\Omega |\hat{T}_{\nu}^{\mu}(x)|\Omega\rangle_\text{ren} &=& \frac{|\alpha|}{960\pi^2}\,\delta_{\nu}^{\mu}\,.
\eeqa
Hence, it vanishes in Minkowski space and non-zero in dS and AdS spaces~\cite{Birrell&Davies,Emelyanov}.
The readings of the detector in Minkowski spacetime explained in terms of the quantum fluctuations fits well, because \eqref{eq:vev_emt_ren} is zero, i.e. no new (Rindler) particles are present~\cite{Sciama&Candelas&Deutsch}.
Although \eqref{eq:vev_emt_ren} does not vanish for de Sitter and anti-de Sitter spaces, the same picture can
be given as well. Indeed, the right-hand side in \eqref{eq:vev_emt_ren} is due to the conformal trace anomaly~\cite{Birrell&Davies}. A geodesic observer in anti-de Sitter space does not register any field excitations, but 
$\langle\Omega |\hat{T}_{\nu}^{\mu}(x)|\Omega\rangle_\text{ren} \neq 0$. On the contrary, a geodesic observer 
in de Sitter space has to register a thermal spectrum with temperature $\text{T}_\text{GH}$, wherein
$\langle\Omega |\hat{T}_{\nu}^{\mu}(x)|\Omega\rangle_\text{ren}$ is the same as in AdS space. Therefore, one 
may conclude that a term due to the particles is absent in \eqref{eq:vev_emt_ren} for dS and AdS spaces as for 
the Minkowski case.

What if one considers the case of the final stage of the collapsing non-rotating matter shell, i.e. the Schwarzschild 
black hole, in the same way? The initial state $|\text{shell}\rangle$ is a coherent state describing the macroscopic system, i.e. the collapsing shell, composed of particles defined with respect to the ordinary vacuum $|\Omega\rangle$. Suppose during the black hole formation, $|\text{shell}\rangle$ evolves unitary in $|\Omega_\text{BH}\rangle$, i.e. the Unruh 
state~\cite{Unruh,Sciama&Candelas&Deutsch,Frolov&Novikov}, and observer's state is unitary equivalent to it. 

From the mathematical point of view, the Hawking effect~\cite{Hawking2} in the case of an eternal black hole can be described in the analogous way as in the above examples~\cite{Sewell,Kay}. That is one separates the local algebra of observables in two mutually independent (commuting) subalgebras with dynamics set by the Killing vector $\partial_t$, where $t$ is the Schwarzschild time. 
Probing the Hartle-Hawking state by local observables belonging to the one of those subalgebras, it appears as a thermal
state with the Hawking temperature.

In the case of the collapsing shell, a similar separation of the local observables has to be realized after the appearance of the
event horizon (like still in a idealized consideration in~\cite{Dappiaggi&Moretti&Pinamonti})~\cite{Emelyanov1}. Thus, the local observables with the help 
of which one can probe the quantum field alters itself, such that a certain part of the field degrees becomes hidden for the observer. 
The observer being in the gravitational field of the black 
hole moves along $\partial_t$ and can register the thermal frequency spectrum by a 
detector~\cite{Sciama&Candelas&Deutsch,Fredenhagen&Haag}. 
However, this situation is slightly different from those treated above, because the renormalized 
$\langle\Omega_\text{BH} |\hat{T}_{\nu}^{\mu}(x)|\Omega_\text{BH}\rangle$ resembles a thermal radiation with the Hawking temperature at the spatial infinity~\cite{Frolov&Novikov}. On the other hand, 
$\langle\Omega_\text{BH} |\hat{T}_{\nu}^{\mu}(x)|\Omega_\text{BH}\rangle$ 
does not look like as for the thermal radiation for an observer being at the finite distance from the black hole. Moreover, 
it is finite on the horizon, whereas for the pure radiation it is divergent as a result of the infinite blueshift of the temperature. Thus, this observer could perhaps similarly interpret the readings of his detector as in the case of the transition from Minkowski space to de Sitter space described above.

\section*{
ACKNOWLEDGMENTS}

It is a pleasure to thank Dr. Alex Vikman and Dr. Michael Haack for valuable discussions during
preparation of this paper. This research is supported by TRR 33 ``The Dark Universe''.

\begin{appendix}
\section{Closed and open coordinates}
\label{app:1}

\paragraph*{Closed coordinates:}
These coordinates $(\eta,\chi,\theta,\phi)$ are related with $x^{\mu}$ as follows
\beqa\nonumber
x^0 &=& \frac{\sin\eta}{\cos\eta + \cos\chi}\,,
\\[0.5mm]
x^1 &=& \frac{\sin\chi}{\cos\eta + \cos\chi}\sin\theta\cos\phi\,,
\\[0.5mm]\nonumber
x^2 &=& \frac{\sin\chi}{\cos\eta + \cos\chi}\sin\theta\sin\phi\,,
\\[0.5mm]\nonumber
x^3 &=& \frac{\sin\chi}{\cos\eta + \cos\chi}\cos\theta\,,
\eeqa
in which the line element \eqref{eq:metric:mink_ds_ads} has the following form
\beqa
ds^2 = \frac{\left(1+|\alpha|\right)^2}{\left((1-\alpha)\cos\chi + (1+\alpha)\cos\eta\right)^2}
\big(d\eta^2 - d\chi^2 - \sin^2\chi d\Omega^2\big)\,.
\eeqa

\paragraph*{Open coordinates:}
These coordinates $(\bar\eta,\bar\chi,\bar\theta,\bar\phi)$ are related with $x^{\mu}$ as follows
\beqa\nonumber
x^0 &=& e^{\bar\eta}\cosh\bar\chi\,,
\\[0.5mm]
x^1 &=& e^{\bar\eta}\sinh\bar\chi\sin\bar\theta\cos\bar\phi\,,
\\[0.5mm]\nonumber
x^2 &=& e^{\bar\eta}\sinh\bar\chi\sin\bar\theta\sin\bar\phi\,,
\\[0.5mm]\nonumber
x^3 &=& e^{\bar\eta}\sinh\bar\chi\cos\bar\theta\,,
\eeqa
in which the line element \eqref{eq:metric:mink_ds_ads} has the following form
\beqa
ds^2 = \frac{\left(1+|\alpha|\right)^2}{\left((1-\alpha)\cosh\bar\eta - (1+\alpha)\sinh\bar\eta\right)^2}
\big(d\bar\eta^2 - d\bar\chi^2 - \sinh^2\bar\chi d\bar\Omega^2\big)\,.
\eeqa

\section{Eigenfunctions of $\mathfrak{I}_\text{C}$}
\label{app:2}

The scalar product between $\Phi_{\mathbf{k}}(x)$ and 
$\mathfrak{I}_\text{C}\Phi_{\mathbf{k}}(x) = x^{-2}\Phi_{\mathbf{k}}(\mathfrak{I}_\text{C}x)$ defines an unitary 
operator $\mathfrak{I}_\text{C}(\mathbf{q}|\mathbf{k})$ on $\mathcal{S}^+$, i.e.
\beqa\nonumber
\mathfrak{I}_\text{C}(\mathbf{q}|\mathbf{k}) &\equiv& (\mathfrak{I}_\text{C}\Phi_{\mathbf{q}},\Phi_{\mathbf{k}}) \;=\; 
-i\int d^3x \left.\big(\mathfrak{I}_\text{C}\Phi_{\mathbf{q}}\partial_0\Phi_{\mathbf{k}}^* -
\Phi_{\mathbf{k}}^*\partial_0\mathfrak{I}_\text{C}\Phi_{\mathbf{q}}\big)\right|_{x^0\;=\;0}
\\[1.5mm]
&=& \frac{-1}{(2\pi)^3}\frac{1}{2(qk)^{\frac{1}{2}}}\int d\mathbf{x} \left(\frac{k}{\mathbf{x}^2} + \frac{q}{\mathbf{x}^4}\right)
e^{i\mathbf{k}\mathbf{x} - i\frac{\mathbf{q}\mathbf{x}}{\mathbf{x}^2}}
\;=\; \frac{-1}{4\pi}\frac{1}{(qk)^{\frac{1}{2}}}\,J_0\big((2q_{\mu}k^{\mu})^{\frac{1}{2}}\big)\,,
\eeqa
where $q \equiv |\mathbf{q}|$ and $k \equiv |\mathbf{k}|$~\cite{Kastrup}.

The operator $\mathfrak{I}_\text{C}$ is unitary and hermitian, hence it has two eigenvalues 
$\pm 1$. The unitary operator that map the modes $\Phi_{klm}(x)$ into the eigenfunctions $\Phi_{nlm}(x)$ of 
$\mathfrak{I}_\text{C}$ is given by
\beqa
E(n|k) &=& 2^{l+1}e^{-\frac{\pi i}{2}(n+l)}\left(\frac{\Gamma\left(1+n\right)}{\Gamma\left(2+2l+n\right)}\right)^{\frac{1}{2}}
k^{l+\frac{1}{2}}e^{-k}L_{n}^{2l+1}(2k)\,, \quad n \;\in\; \mathbb{N}_0\,,
\eeqa
such that
\beqa
\Phi_{nlm}(x) &=& \int dk E^*(n|k)\Phi_{klm}(x)\,,
\eeqa
One further obtains
\beqa
\Phi_{nlm}(x) &=& \int\limits_{-\infty}^{+\infty}dp\;\Phi_{plm}(x)
\int\limits_{0}^{+\infty}dk\;E^*(n|k)C(p|k) \;\equiv\; \int\limits_{-\infty}^{+\infty}dp\;M^*(n|p)\Phi_{plm}(x)\,,
\eeqa
where $M(n|p)$ is given in \eqref{eq:M}, so that one has
\bsubeqs
\beqa
(\Phi_{plm},\Phi_{nl^{\prime}m^{\prime}})_{\Sigma} &=& M(n|p)\delta_{ll^{\prime}}\delta_{mm^{\prime}}\,,
\\[1mm]
(\Phi_{plm},\Phi_{nl^{\prime}m^{\prime}}^*)_{\Sigma} &=& 0\,.
\eeqa
\esubeqs

\section{Relating $\Phi_{nlm}(x)$ and $\Phi_{plm}(x)$}
\label{app:3}

One has to calculate the product of $\Phi_{plm}(x)$ given by \eqref{eq:phi_plm} and $\Phi_{nlm}(x)$ given by \eqref{eq:phi_nlm}:
\beqa\nonumber
(\Phi_{plm},\Phi_{nl^{\prime}m^{\prime}})_{\Sigma} &=&
(\Phi_{plm},\Phi_{nl^{\prime}m^{\prime}})_{\Sigma_1\cup\Sigma_2}
\\[1.5mm]
&=& (\Phi_{plm},\Phi_{nl^{\prime}m^{\prime}})_{\Sigma_1} + 
(\mathfrak{I}_\text{C}\Phi_{plm},\mathfrak{I}_\text{C}\Phi_{nl^{\prime}m^{\prime}})_{\Sigma_2}\,.
\eeqa
Substituting the modes, one obtains
\beqa
(\Phi_{plm},\Phi_{nl^{\prime}m^{\prime}})_{\Sigma_1} &=&A_{nl}^*B_{pl} \frac{i^{n+l+1}e^{+\frac{\pi p}{2}}}{\Gamma\left(l+\frac{3}{2}\right)}
\times
\left\{
\begin{array}{ll}
\frac{2(-1)^mJ_1^{+}(2m+1,l)}{B\left(1+l,1+m\right)}\,, & n \;=\; 2m+1\,, 
\\[3mm]
\frac{(-1)^mJ_2^{+}(2m,l)}{(1+l+m)B\left(1+l,1+m\right)}\,, & n \;=\; 2m\,,
\end{array}
\right.
\eeqa
where $m \in \mathbb{N}_0$, and
\beqa
(\mathfrak{I}_\text{C}\Phi_{plm},\mathfrak{I}_\text{C}\Phi_{nl^{\prime}m^{\prime}})_{\Sigma_2} &=& 
A_{nl}^*B_{pl} \frac{i^{n+l+1}e^{-\frac{\pi p}{2}}}{\Gamma\left(l+\frac{3}{2}\right)}
\times
\left\{
\begin{array}{ll}
\frac{2(-1)^mJ_1^{-}(2m+1,l)}{B\left(1+l,1+m\right)}\,, & n \;=\; 2m+1\,, 
\\[3mm]
\frac{(-1)^{m+1}J_2^{-}(2m,l)}{(1+l+m)B\left(1+l,1+m\right)}\,, & n \;=\; 2m\,,
\end{array}
\right.
\eeqa
where
\beqa
A_{nl} &=& \frac{i^{n+l+1}}{2^{l+1}}\frac{\Gamma\left(2+2l\right)}{\Gamma\left(l+\frac{3}{2}\right)}
\left(\frac{\Gamma\left(1+n\right)}{\Gamma\left(2+2l+n\right)}\right)^{\frac{1}{2}}, \quad
B_{pl} \;=\; \frac{i^l}{(4\pi)^{\frac{1}{2}}}\Gamma\left(1+l+ip\right)\,
\eeqa
and
\bsubeqs
\beqa
J_1^{\pm}(n,l) &\equiv& \int\limits_{0}^{+\infty}dx\left(p \pm \frac{n+l+1}{\cosh x}\right)
\frac{\left(\sinh x\right)^{l+\frac{3}{2}}}{\left(\cosh x\right)^{l+2}}\left(\tanh \frac{x}{2}\right)^{l+\frac{1}{2}}
\\[1.5mm]\nonumber
&&\times\;_2F_1\left(\frac{1}{2} - ip, \frac{1}{2} + ip, l +\frac{3}{2}; -\sinh^2\frac{x}{2}\right)\;
_2F_1\left(\frac{1-n}{2},\frac{3}{2} + l + \frac{n}{2},\frac{3}{2}; \frac{1}{\cosh^2x}\right),
\\[1mm]
J_2^{\pm}(n,l) &\equiv& \int\limits_{0}^{+\infty}dx\left(p \pm \frac{n+l+1}{\cosh x}\right)
\frac{\left(\sinh x\right)^{l+\frac{3}{2}}}{\left(\cosh x\right)^{l+1}}\left(\tanh \frac{x}{2}\right)^{l+\frac{1}{2}}
\\[1.5mm]\nonumber
&&\times\;_2F_1\left(\frac{1}{2} - ip, \frac{1}{2} + ip, l +\frac{3}{2}; -\sinh^2\frac{x}{2}\right)\;
_2F_1\left(\frac{-n}{2},1 + l + \frac{n}{2},\frac{1}{2}; \frac{1}{\cosh^2x}\right).
\eeqa
\esubeqs

One can show that these integrals satisfy the following recurrence equations
\bsubeqs\label{eq:recurrence_eq}
\beqa
J_1^{\pm}(n,l+1) &=& \frac{2l + 3}{4\left(p^2 + (1+l)^2\right)}
\Big((2+2l+n)J_1^{\pm}(n,l) - (2+n)J_1^{\pm}(n+2,l) \Big)\,,
\\[1mm]
J_2^{\pm}(n,l+1) &=& \frac{2l + 3}{4\left(p^2 + (1+l)^2\right)}
\Big((3+2l+n)J_2^{\pm}(n,l) - (1+n)J_2^{\pm}(n+2,l) \Big)\,
\eeqa
\esubeqs
and can be exactly evaluated for $l = 0$~\cite{Emelyanov}, so that
\bsubeqs
\beqa
J_1^{\pm}(n,0) &=& \mp \frac{\pi^2\exp\left(\pm\frac{\pi p}{2}\right)}{2^{\frac{3}{2}}\sinh^2\pi p}
\frac{_2F_1\left(-1-n,ip,ip-n;-1\right)}{\Gamma\left(2+n\right)\Gamma\left(1-ip\right)\Gamma\left(ip-n\right)}\,,
\\[1mm]
J_2^{\pm}(n,0) &=& -i\frac{\pi^2\exp\left(\pm\frac{\pi p}{2}\right)}{2^{\frac{3}{2}}\sinh^2\pi p}
\frac{_2F_1\left(-1-n,ip,ip-n;-1\right)}{\Gamma\left(1+n\right)\Gamma\left(1-ip\right)\Gamma\left(ip-n\right)}\,.
\eeqa
\esubeqs
One can further check that the recurrence Eq.~\eqref{eq:recurrence_eq} are solved by
\bsubeqs
\beqa\nonumber
J_1^{\pm}(n,l) &=& \mp(-1)^n\frac{i\pi\exp\left(\pm\frac{\pi p}{2}\right)}{2^{n+\frac{1}{2}}\sinh(\pi p)}
\frac{\Gamma\left(1+l+n-ip\right)}{\Gamma\left(1+l-ip\right)}
\frac{\Gamma\left(l+\frac{3}{2}\right)}{\Gamma\left(\frac{2+n}{2}\right)\Gamma\left(l + \frac{3+n}{2}\right)}
\\[1mm]
&&\times\; _2F_1\left(-n,1+l+ip,-l-n+ip;-1\right)\,,
\\[4.00mm]\nonumber
J_2^{\pm}(n,l) &=& (-1)^n\frac{\pi\exp\left(\pm\frac{\pi p}{2}\right)}{2^{n+\frac{1}{2}}\sinh(\pi p)}
\frac{\Gamma\left(1+l+n-ip\right)}{\Gamma\left(1+l-ip\right)}
\frac{(1+n)\Gamma\left(l+\frac{3}{2}\right)}{\Gamma\left(\frac{3+n}{2}\right)\Gamma\left(l + \frac{2+n}{2}\right)}
\\[1mm]
&&\times\; _2F_1\left(-n,1+l+ip,-l-n+ip;-1\right)\,.
\eeqa
\esubeqs
Thus, one finds
\bsubeqs
\beqa
(\Phi_{plm},\Phi_{nl^{\prime}m^{\prime}})_{\Sigma} &=& M(n|p)\delta_{ll^{\prime}}\delta_{mm^{\prime}}\,,
\\[1mm]
(\Phi_{plm},\Phi_{nl^{\prime}m^{\prime}}^*)_{\Sigma} &=& 0\,.
\eeqa
\esubeqs

\end{appendix}

\end{document}